\documentclass[12pt,preprint]{aastex}

\newcommand{\ximax}{\xi_{\mathrm{max}}}

\shorttitle{Inner Structure of L694--2}
\shortauthors{Harvey et al.}
\begin{document}

\title{Inner Structure of Starless Core L694--2 Derived from 
Millimeter-Wave Interferometry\footnote{Based on observations 
carried out with the IRAM Plateau de Bure Interferometer and the Berkeley 
Illinois Maryland Array. IRAM is supported by INSU/CNRS (France), MPG 
(Germany) and IGN (Spain). The BIMA array is operated
by the Berkeley-Illinois-Maryland Association under funding from the
National Science Foundation.}}

\author{Daniel W.A.\ Harvey,
        David J.\ Wilner, 
        Philip C.\ Myers}

\email{dharvey, dwilner, pmyers@cfa.harvard.edu}

\affil{Harvard-Smithsonian Center for Astrophysics, 60 Garden Street,
Cambridge, MA 02138}

\and

\author{Mario Tafalla}
\email{m.tafalla@oan.es}
\affil{Observatorio Astron\'{o}mico Nacional, Alfonso XII 3, E-28014
Madrid, Spain}

\begin{abstract}
We study the density structure of the candidate contracting starless core 
L694--2 using 1.3~mm dust continuum observations from the IRAM Plateau de 
Bure Interferometer and the Berkeley-Illinois-Maryland Array, which probe
spatial scales from 10000~AU to 500~AU.
The long baseline PdBI observations detect no emission from the core, 
and limit the maximum contamination from a compact component 
$F_c < 2.7$~mJy. The flux limit corresponds to a very small disk mass, 
$M_{\mathrm{disk}} \lesssim 5 \times 10^{-4}$~M$_{\odot} \, 
(60$~K${}/T_{\mathrm{disk}})$, and bolsters the ``starless'' interpretation
of the L694--2 core. The shorter baseline BIMA data are 
compared to a series of density models using a physically motivated 
temperature distribution with a central minimum.  
This analysis provides clear evidence for a turn-over from the steep 
density profile observed in the outer regions in dust extinction 
to substantially more shallow behavior in the inner regions ($< 7500$~AU). 
The best fit Bonnor-Ebert, Plummer-like, broken power law, and 
end-on cylinder models produce very similar flattened profiles
and cannot be distinguished. We quantify the sensitivity 
of the inferred structure to various uncertainties, including the 
temperature distribution, the accuracy of the central position, and 
the presence of a weak unresolved central component. The largest uncertainty 
comes from the temperature assumption; an isothermal model modifies 
the best fit parameters by $\sim 2\, \sigma$, with the inferred density 
profiles more shallow. Dust emission and extinction profiles are 
reproduced by an embedded isothermal cylinder with scale height $H=13.5''$ 
inclined at a small angle to the line of sight. The turn-over observed in 
the L694--2 density distribution suggests that pressure forces still 
support the core, and that it has not fully relaxed as in the 
inside-out collapse model, despite the extended inward motions inferred 
from molecular line observations (Lee, Myers, \& Tafalla 2001). 
In the context of the cylindrical density model, these inward motions 
may represent the contraction of a prolate core along its major axis.
\end{abstract}


\section{Introduction}
The observed properties of dense cores (e.g.\ Benson \& Myers 1989) 
form the basis of the standard model of isolated star formation. 
In this model, the ``starless'' dense core represents the earliest 
identifiable stage of the star formation process. The physical conditions 
in this early stage have a profound impact on the evolution of protostars 
towards the main sequence. The initial density structure, particularly 
in the innermost regions, affects the collapse dynamics and the 
time dependence of the mass accretion rate and therefore many of 
the observable properties of protostars, including luminosity.

A quantitative understanding of the collapse dynamics has been hindered 
by uncertain knowledge of appropriate initial conditions (Andr\'{e},
Ward-Thompson \& Barsony 2000). In the popular theory of ``inside-out'' 
collapse, a spherical starless core loses turbulent and magnetic support 
and relaxes to a balance between gravity and thermal pressure, an $r^{-2}$ 
density distribution is established and the core collapses from the 
inside-out with a constant mass accretion rate (Shu 1977). However,
if collapse begins before the density distribution fully relaxes, then a 
central region of relatively constant density remains and the mass accretion
rate is an order of magnitude larger at early times (Foster \& Chevalier 
1993). This phenomenon has been identified with the youngest ``Class~0'' 
protostars, which exhibit especially powerful outflows (Henriksen, Andre 
\& Bontemps 1997; Andre, Ward-Thompson \& Barsony 2000). Better observations
of starless cores are needed to determine the initial conditions.

The L694--2 dense core is one of several ``strong'' infall candidates among
starless cores, based on observations of molecular line profiles with
redshifted self-absorption in a systematic search of more than 200 targets
(Lee, Myers \& Tafalla 1999). Molecular line maps of these objects provide 
strong evidence of inward motions, with speed $\sim 0.1$~km~s$^{-1}$ over 
a radius of $\sim 0.1$~pc (Lee, Myers \& Tafalla 2001). The physical basis
for these motions is unclear. The speeds are subsonic and may be associated
with condensation through ambipolar diffusion, or perhaps a magnetically
diluted gravitational collapse (Ciolek \& Basu 2000). Alternatively, 
pressure driven motions due to the dissipation of turbulence may be 
responsible (Myers \& Lazarian 1998). 

Observations of dust column density provide a means to infer the 
structure and dynamical state of a dense core. 
Harvey et al.\ (2003b) describe near-infrared extinction measurements 
toward L694--2 that show a very steep density profile in the region 
$r=30''$ to $83''$ (0.036 to 0.1~pc, or 7500 to 20000~AU), where 
background stars were detectable for extinction measurements. 
The extinction data can be fit by a number of models that produce nearly 
degenerate column density profiles, including a simple power law with 
index $p=2.6 \pm 0.2$, a supercritical Bonnor-Ebert sphere with 
dimensionless outer radius $\ximax=25 \pm 3$, or a nearly end-on
isothermal cylinder with scale height $H=13.5'' \pm 1.5''$. The latter 
two models provide a physical description of the core that represents an
unstable configuration of material that is consistent with the infalling 
motions inferred from molecular spectral line profiles. In particular, the
slightly tilted cylindrical model that provides a basis for interpreting 
the asymmetry of the L694--2 core should be highly unstable to gravitational 
collapse along its axis (close to the line of sight), consistent with the
strength and orientation of the observed velocity structure. The model also
provides a framework for understanding the very high extinctions observed 
in L694--2, namely that the core has prolate structure with the full extent 
(mass) of the core being missed by analysis that assumes symmetry between 
the profile of the core in the plane of the sky, and the profile of the core
along the line-of-sight.

Observations of long wavelength dust emission provide another means
of probing density structure. This technique is almost as direct as 
observations of dust extinction. The intensity of the optically thin
emission provides an integral along the line-of-sight of the product 
of the density, temperature, and opacity of the dust. 
The analysis of dust emission complements dust extinction work because 
the dust emission method becomes most effective in the high column density 
inner regions of dense cores where dust extinction becomes large and 
difficult to penetrate. 

Much of the detailed information on starless core structure comes from 
observations of dust emission, using data from bolometer cameras 
(e.g. Ward-Thompson, Motte \& Andre 1999,
Shirley et al. 2000, Visser, Richer \& Chandler 2001).
An important conclusion from these studies is that starless cores 
appear to show flat density profiles in their inner regions, 
with extended envelopes that fall off rapidly in power law fashion.
Recently, the predominance of the flat central density gradients
has been called into question as more sophisticated analysis including 
self-consistent temperature calculations indicate much smaller regions of 
flattening, or no flattening at all, in large part because the cores are   
cooler in their deep interiors than assumed previously (Evans et al.\ 2001,
Zucconi et al.\ 2001). In addition, the central regions of even the 
nearest protostellar cores are generally comparable in size to the 
beamwidths of these telescopes and are poorly resolved. 
The density structure at smaller scales can be probed with interferometers. 
Strong constraints can be derived by analyzing interferometer data from
dense cores directly in the visibility domain, though this approach 
has been rarely used (Keene \& Masson 1990, Hogerheijde et al.\ 1999, 
Harvey et al.\ 2003a, Looney, Mundy \& Welch 2003).

L694--2 has been mapped at 850~$\mu$m with SCUBA on the JCMT by Visser 
(2000) and at 1.2~mm with the IRAM 30m by Tafalla et al.\ (2003). 
The two maps have similar resolutions, 14'' for SCUBA and 11'' for IRAM. 
The dust emission profiles suggest a steep outer density gradient 
($p\simeq 2.7$ from SCUBA, $p=2.5$ from IRAM) consistent with
that inferred from dust extinction (Harvey et al.\ 2003b). The usual 
isothermal analysis suggest a flattening of the density gradient within a 
radius of a few thousand AU. Visser (2000) fits a broken power law model
that suggests an inner power law index of $p=0.8$ within a break radius 
of 8000~AU ($\sim 30''$). Although no radio point source has been detected 
in the L694--2 core (e.g.\ Harvey et al.\ 2002), the measurement 
of the inner power law index suffers a potentially large systematic error 
from possible point-source contamination to the central beam (Shirley et 
al.\ 2002). In this paper, we present observations of dust continuum 
emission from L694--2 at 1.3~mm obtained with the IRAM Plateau de Bure 
Interferometer (PdBI) and the Berkeley-Illinois-Maryland-Array (BIMA). 
The long baseline observations from the PdBI limit the contribution to the 
flux from any embedded point source, while the BIMA observations sample 
the dense core structure on size scales from $\sim 10000$~AU to 
$\sim 1500$~AU 
(resolution to $6''$). Together, these observations provide conclusive 
evidence of a flattening in the density profile within $\sim 30''$ from the 
center of the core.

\section{Observations} 

\subsection{IRAM PdBI}
Continuum emission from L694--2 was observed at 1.3~mm (231.32~GHz) with 
the IRAM PdBI in the compact D configuration on 2001 November 20 (4 antennas)
and 2002 April 03 (6 antennas). Table~1 lists the observational parameters.
A single pointing was used, with center 19:41:04.44, 10:57:00.9 (J2000) 
chosen to coincide with the position of peak emission in the 
N$_2$H$^+$(1--0) spectral line measured at BIMA (Jonathan Williams, private 
communication). The half-power field of view for PdBI at this wavelength is 
$22''$ (5500~AU). The PdBI observations provide measurements at baseline 
lengths from the shadowing limit of 15~m to a maximum of over 100~m 
(resolution $\sim 2\farcs6$). The absolute flux scale was set by 
observations of the standard source MWC~349, assumed to be 1.71~Jy.
The estimated uncertainty in the flux scale is roughly 20\%. Frequent 
observations of nearby calibrators J1751+096 and J1925+211 were used to 
determine time-dependent complex gains. Continuum visibility records were 
formed for each 60~s integration of the digital correlator 
($3\times 160$~MHz bandwidth, tuning double sideband). The correlator 
bandpass was measured with observations of the strong sources 3C~345 (2001 
November 20) and 3C~273 (2002 April 03). The data were calibrated using the 
IRAM software package {\em CLIC}, and comprise a total of 7240 records (2040 
records from 2001 November 20, 5200 from 2002 April 03). In addition to 
amplitude and phase, each record contains a variance measure, determined 
from the system temperature and antenna gains. 

\subsection{BIMA}
Continuum emission from L694--2 was observed at 1.3~mm (231.32~GHz) with
BIMA in the compact D configuration on 2001 September 22 and 29, 2002 June 
08, and 2002 September 09, and in the more extended C configuration on 
2002 March 31. Table~1 lists the observational parameters.
A single pointing of the nine antennas was used on each occasion. 
The pointing center for the BIMA observations is displaced from the 
PdBI pointing, by $\delta$R.A.$=-1\farcs8$, $\delta$Dec.$=5\farcs2$. 
The half-power field of 
view for the BIMA antennas is $50''$ (12,500~AU). The BIMA observations 
provide measurements at baseline lengths from the shadowing limit of 6~m to 
a maximum of 70~m (resolution $\sim 3\farcs7$). The bandpass and flux 
calibration were determined through observations of Uranus, using the 
a~priori antenna gains. The estimated uncertainty in the flux scale is 
roughly 15\%. Frequent observations of J1925+211 were used to determine 
time-dependent complex gains. Continuum visibility records were formed for 
each 23~s integration of the digital correlator (700~MHz bandwidth). 
The data were calibrated in the BIMA software package {\em MIRIAD}. 
The resulting $\sim 2 \times 10^5$ records were then averaged in time bins
of 3.45~minutes (9 records). This choice of binning was made to reduce 
the number of visibilities to a more manageable size for analysis, without 
introducing significant phase error at the longer baselines. The resulting 
dataset contains 22327 records. As with the PdBI dataset, a variance measure
for each visibility measurement is also recorded. 

\section{Constructing Model Visibilities}
The visibility measurements are analyzed directly, without producing images
that are limited by standard Fourier inversion and deconvolution techniques.
This approach is computationally intensive, but it allows a much more direct 
comparison with models than analyzing images. In particular, the results 
are not compromised by problems with the synthesized beam characteristics 
and dynamic range.

The L694--2 visibilities are compared to theoretical models of protostellar  
envelope structure by constructing synthetic visibilities, taking account 
of (1) the dust continuum radiative transfer, and (2) the specifics of 
the observations, including the exact $(u,v)$ sampling and primary beam
pointing and attenuation for the two telescopes. The models necessarily
include an assumed temperature distribution and (constant) specific mass 
opacity in addition to the model density field. We do not consider an 
exhaustive list of starless core models but instead fix attention on 
a few widely promoted density fields, including Bonnor-Ebert spheres, 
Plummer-like models, broken power law descriptions, and isothermal 
cylinders. These models can all match the steep density gradient 
inferred for the outer regions of the core.

The model datasets are constructed using the recipe described in Harvey et
al.\ (2003a). In brief, a $512\times 512$ model intensity image of resolution
$0\farcs5$~pixel${}^{-1}$ is calculated using the full Planck function for
the emissivity and integrating the radiative transfer equation through the
model globule. Each model is normalized to a flux at 1.2~mm of 
$800 \pm 80$~mJy within a circular aperture (top-hat) of radius $30''$, 
calculated from the continuum map of L694--2 made by Tafalla et al.\ (2003)
with MAMBO on the IRAM 30m (assuming a dust opacity spectral index of unity).
We adopt an outer boundary of $R_{\mathrm{out}}=0.15$~pc in the models based
on the extinction observations (Harvey et al.\ 2003b), although the results
are not sensitive to this assumption. Observations are simulated by
performing an FFT, and assuming a Gaussian form for the primary beams 
(PdBI FWHM $22''$, BIMA FWHM $50''$). The exact $(u,v)$ sampling is achieved
by interpolating the real and imaginary parts of the resulting visibility 
grid. The center for each model is assumed to be at the pointing center of
the PdBI observations, and in Section~\ref{sec:uncer} we discuss the slight 
sensitivity of the results to this assumption. 

\subsection{Model Selections}
\label{sec:models}
As detailed above, the model visibilities are derived from the assumed
(1) mass density distribution $\rho(r)$, (2) dust temperature distribution 
$T_d(r)$, and (3) specific mass opacity of the dust 
$\kappa_{1.3~\mathrm{mm}}$
(normalized to a constant flux within a $30''$ circular aperture). 
We consider the expected form of these quantities, and their uncertainties.

\subsubsection{Density}
The main goal is to constrain the density distribution of L694--2, 
given a realistic choice for the temperature distribution, and specific
mass opacity. We consider four models of starless core density structure 
that can match the steep density gradient in the outer regions inferred 
from the near-infrared extinction.

{\em Bonnor-Ebert spheres}.--- These models are pressure-confined isothermal
spheres, for which the solution remains finite at the origin (Ebert 1955,
Bonnor 1956). They are solutions of a modified Lane-Emden equation
(Chandrasekhar 1967):
\begin{equation}
\frac{1}{\xi^2} \frac{d}{d \xi} \left( \xi^2 \frac{d \psi}{d \xi}
\right) = \exp{(-\psi)} \ ,
\end{equation}
where $\xi=(r/R_0)$ is the dimensionless-radius,  
$R_0=a/\sqrt{4 \pi G \rho_c}$ is the (physical) scale-radius, and 
$\psi( \xi)=- \ln{ (\rho/ \rho_c )}$ is a logarithmic density contrast, with
$\rho_c$ the (finite) central density, and $a$ is the effective sound speed
in the core (we adopt $a=0.20$~km~s$^{-1}$ based on an assumed central
temperature of 9~K, with turbulent component 
$a_{\mathrm{turb}}=0.09$~km~s$^{-1}$, as in the extinction study). 
Configurations with $\ximax > 6.5$ are unstable to gravitational collapse. 
A highly supercritical dimensionless-radius of $\ximax=25 \pm 3$ was found 
to reproduce the extinction observations of L694--2 for radii $r \geq 30''$.

{\em Plummer-like models.}--- These are empirical models suggested by
Whitworth \& Ward-Thompson (2001) that capture the essential observed
properties of starless cores with a minimum of free parameters. The 
model assumes that when a prestellar core becomes unstable against 
collapse at time t=0, it is static and approximates to a Plummer-like 
density profile (Plummer 1911), of the form:
\begin{equation}
\rho(r, t=0)=\rho_0 \left[ 
\frac{R_0}{(R_0^2+r^2)^{1/2}} \right]^{\eta}
\end{equation}
The initial density is therefore uniform for $r \ll R_0$, and falls off 
as $r^{-\eta}$ for $r \gg R_0$. Whitworth \& Ward-Thompson (2001) propose 
a fixed value of $\eta=4$ in the model in order to reproduce the relative 
lifetimes and accretion rates for the Class~0 and Class~I phases. For 
L694--2, the density power law index in the outer regions inferred from 
near infrared extinction is $p=2.6 \pm 0.2$ if the core is assumed to be 
unobscured, but has a steeper value of $p=3.7 \pm 0.3$ if the core is 
embedded in a more extended uniform distribution of material, as is 
suggested by the flattening in the radial profile beyond $r \gtrsim 0.1$~pc 
(Harvey et al.\ 2003b). A Plummer-like core with $\eta=4$ embedded in a 
more extended cloud can therefore reproduce the extinction observations. 
The scale radius $R_0$ is not constrained by the extinction measurements 
because no flattening is evident in the data (which do not penetrate the 
$r \lesssim 30''$ inner region).
 
{\em Broken Power Law models.}--- These models represent a variation of the
Plummer-like model, to allow for an inner density power law index that
is non-zero. The density distribution we adopt is of the form:
\begin{equation}
\rho(r) = \frac{\rho_0}{((r/R_0)^{(2 p)} + (r/R_0)^{(2 \eta)})^{(1/2)}}
\end{equation}
For $r \ll R_0$, the density falls off as $r^{-p}$, and for $r \gg R_0$, 
the density falls off as $r^{-\eta}$. We choose this continuous prescription
to prevent discontinuities in the first derivatives (which show up in the 
visibility profile) that would result from including a unphysical sharp
break in the density distribution. We choose $\eta=2.6$ to reproduce the
observed extinction without introducing additional structure components. 
There is some degeneracy between $R_0$ and $p$ in the resulting visibility
(or intensity) profile, since decreasing $R_0$ or increasing $p$ both 
produce a steeper emission profile or a flatter visibility profile. The
visibility amplitude for a typical starless core profile falls rapidly and 
is already very low at baselines that are long enough to sample the inner 
structure that cannot be probed with extinction. Even the high quality
visibility dataset that we have obtained from BIMA and the PdBI therefore 
does not provide sufficient sensitivity to constrain both
parameters simultaneously. We therefore choose a fixed value of $R_0=30''$ 
in order to quantify the effect of the inner power law index $p$. This choice
is made on the basis that $R_0=30''$ is the largest value of $R_0$ that is
consistent with the extinction data, which will lead to the largest possible
value of $p$ allowed by the dataset (since a smaller value of $R_0$ will
necessarily require a shallower $p$ to produce a given visibility amplitude
at a given baseline). 

{\em Cylinder models.}--- The isothermal cylinder model is a two dimensional
analog of the Bonnor-Ebert sphere. The density is a function of the radial
coordinate only (Ostriker 1964):
\begin{equation}
\rho(r, z) = \frac{\rho_c}{(1 + ( r^2 / 8 H^2))^2}
\end{equation}
where $H=a/\sqrt{4 \pi G \rho_c}$ is the scale height that is equivalent to
the scale radius $R_0$ in the B-E analysis. The density is uniform near the
axis of the cylinder but decays ever more rapidly with increasing radius,
asymptoting to a power law of index $p=4$ for $r \gg H$.
The filament is supported radially by pressure gradients, but is unstable 
in the direction along its axis. The density distribution of the isothermal 
cylinder is a two dimensional case of the Plummer-like model, with 
$\sqrt{8} H = R_0$, and with a physical basis for the 
normalization of the density profile. A cylindrical model can therefore 
reproduce the observational properties of pre-stellar cores that provided the
motivation for the Plummer-like models. In addition, because the 
isothermal cylinder and the B-E sphere both represent equilibria between 
self-gravity and gas pressure, the spherically averaged density profile of 
an isothermal cylinder can also mimic closely that of a B-E sphere, in 
particular a flat inner region with a steeply falling envelope (Boss \& 
Hartmann 2001). The cylindrical model also provides a basis for interpreting
the departures from spherical symmetry in the L694--2 core (Harvey et al.\ 
2003b). A slightly tilted, embedded cylinder with scale height 
$H=13.5''\pm 1.5''$ reproduces the extinction profile for the inner $83''$ 
(0.1~pc) of the core. In the present study we consider isothermal cylinders
viewed along the axis, since the subtle effect of a small tilt angle can not
be constrained with the visibility dataset.

\subsubsection{Extended Cloud Structure}
Both Plummer-like and Isothermal Cylinder models can successfully describe
the near-IR extinction observations, if the cores are embedded in
an extended distribution of gas. This additional component to the density
structure is suggested by the shape of the radial extinction profile, and 
must be accounted for in the fitting of these types of model. We include the
extended cloud structure in the visibility analysis as follows. The extended 
structure is assumed to be smooth and is therefore resolved out by the 
interferometers. The only effect of the additional structure is to reduce 
the total flux that is attributable to the core in this context. 
The extinction profile of the L694--2 core asymptotes to a color-excess that
is approximately 0.2 magnitudes higher than the background, and about a 
tenth ($1/10$) of the color excess at $30''$ radius (the edge of the flux 
normalization aperture). We approximate the intensity profile to be flat 
within $30''$ of the core, and assume that the temperature of the extended 
gas is equal to that in the core, so that the extended structure accounts 
for 10\% of the flux normalization for each of these types of model.

\subsubsection{Temperature}
A detailed study of the expected dust temperature distribution in starless 
cores has been performed by Evans et al.\ (2001). They calculate the 
temperature distribution, $T_d(r)$, self-consistently using a 1D radiative
transport code, and assuming Ossenkopf \& Henning (1994) opacities for 
grains that have grown by coagulation and accretion of thin ice mantles. 
They find that the Interstellar Radiation Field (ISRF) dominates the heating
in the core, roughly a factor of 3 stronger than heating due to cosmic rays, 
even at the center of an opaque core. The effect of the opacity law on 
$T_d(r)$ is small; the largest difference is at the center of the core, 
where opacities for coagulated grains that lack mantles cause a lower value 
of $T_d$ by $\sim 0.5$~K.

Evans et al.\ (2001) present the dust temperature distribution $T_d(r)$ for 
a Bonnor-Ebert sphere with outer radius $R_{\mathrm{out}}=0.17$~pc (35000~AU)
in which the gas is isothermal at 10~K (their Figure~3). The model has a 
central density of $n=1\times 10^6$~cm$^{-3}$, and a dimensionless outer 
radius of $\ximax\simeq 42$, and is heated by an ISRF that combines the 
infrared behavior from COBE (Black 1994) and the ultraviolet behavior from
Draine (1978). The dust temperature varies from a minimum of $\sim 8$~K in 
the inner thousand AU of the core, to a maximum of $\sim 14$~K in the outer 
regions. The model is similar in radius, but somewhat more centrally 
condensed than the B-E sphere that best fits the extinction in L694--2
($R_{\mathrm{out}}=0.17$~pc, $R_0=3.4''$ in the computed model; 
$R_{\mathrm{out}}=0.15 \pm 0.14$~pc, $R_0=4.9'' \pm 0.6''$ for L694--2).
In addition, the Evans et al.\ (2001) analysis neglects any shielding by an
extended component of material that surrounds the core, as is seen around
L694--2 in near-infrared extinction (Harvey et al.\ 2003b). The degree to 
which the interior of the L694--2 globule is shielded from the ISRF will 
differ somewhat from in the Evans et al.\ (2001) model. However, the Evans 
et al.\ (2001) model likely provides an accurate description of the actual
temperature profile in L694--2 to within the uncertainties that stem from 
the inhomogeneities of the ISRF. We therefore adopt the Evans 
et al.\ (2001) temperature distribution for use in our modeling. We 
investigate the sensitivity to the assumed temperature distribution by also
exploring models that assume a constant dust temperature of $T_d=12$~K.

The Bonnor-Ebert and Cylinder models used in the model fitting are based 
on hydrostatic equilibrium density configurations for a gas cloud that 
has an isothermal kinetic temperature $T_K$. The Evans et al.\ (2001)
temperature distribution is calculated using a density model that obeys
the same assumption. At high densities efficient gas-dust coupling 
forces $T_K$ to equal the dust temperature $T_d$, while at low densities
($n \lesssim 10^4$~cm$^{-3}$) $T_K \neq T_d$ (Takahashi, Silk, \& 
Hollenbach 1983, Doty \& Neufeld 1997). The Bonnor-Ebert model that best 
fits the L694--2 extinction has central density  
$n\sim 3 \times 10^5$~cm$^{-3}$, with a center-to-edge density contrast of 
$\sim 400$. In the inner regions of L694--2 the kinetic temperature of the 
gas should decrease in a similar fashion to the dust temperature, but in the
outer regions the two temperature distributions may depart.
An entirely self-consistent approach to the problem would include a full 
calculation of the gas energetics, including dust coupling to calculate
simultaneous density and (dust) temperature distributions. Such an approach
has been followed by Galli, Walmsley \& Gon\c{c}alves (2002). However,
to zeroth order, small departures from isothermality in the kinetic 
temperature have little effect on the density profile, since the pressure 
gradient of the gas is dominated by the gradient in the density and not the 
temperature. The effect of non-isothermality on the resulting emission 
profile is therefore dominated by the dust temperature dependence of the 
Planck function. Since the uncertainties that surround the calculation of 
the dust temperature profile are substantial, the computational overhead of 
a more self-consistent approach is not justified at this time. We separate 
the problem into two parts, modeling density profiles assuming a constant 
gas kinetic temperature, but relaxing the isothermal assumption in order to 
model the resulting thermal emission from the core.

A remaining issue deals with the appropriate temperature distribution to 
use for the cylindrical models, since these are by definition not 
spherically symmetric. Since no self-consistent 3D dust radiative transfer 
results are available for this type of model, we have made a very simple 
approximation, assuming the filament to have an aspect ratio of 2:1 and 
stretching the Evans et al.\ (2001) temperature distribution appropriately 
along the line-of-sight.

\subsubsection{Mass Opacity}
The mass opacity of dust grains in the millimeter region of the spectrum in
protostellar envelopes is uncertain but generally assumed to follow a
power law with frequency, $\kappa_{\nu} \propto \nu^{\beta}$. The power-law
index varies depending on the dust properties, but tends to be bounded by a
small range, roughly 1 to 2 (Ossenkopf \& Henning 1994). The opacity of the
dust affects the temperature distribution of the dust, as already described.
For our purposes, only the spectral index is of importance, since we assume 
a constant opacity (independent of $r$), and the overall normalization is 
fixed by matching the MAMBO single dish observation. The spectral index 
affects the normalization of the models only weakly since the MAMBO flux 
constraint is made at a wavelength very close to that of the PdBI and BIMA 
observations. We have adopted $\beta=1$; the extreme alternative of 
$\beta=2$ leads to only a 4\% change in flux normalization (for $T_d=12$~K),
which is small compared to the overall 10\% uncertainty in the normalization
itself.

\section{Method for Fitting Model Parameters and Evaluating Fit Quality}
The basic procedure is to maximize the probability distribution:
\begin{equation}
 P(\mbox{Model} \: | \: \mbox{data})=
\prod_i{e^{-(Z_i-f(x_i; p, m))^2/2\sigma_i^2}} \,
e^{-(m - m_0)^2/2\sigma_m^2}
\end{equation}
where the $Z_i$ are the visibility data points with uncertainty $\sigma_i$,
$f(x_i; p, m)$ are the model data points, $p$ a free parameter in the
models, and $m$ a model parameter about which we have a constraint (namely
that it is a Gaussian random variable with mean $m_0$ and standard deviation
$\sigma_m$). Maximizing the probability distribution is equivalent to
minimizing the logarithm of its inverse. Taking account the fact that the
$Z_i$ are by nature complex visibilities, we want to minimize a modified
$\chi^2$:
\begin{equation}
\tilde{\chi}^2 = \sum_i{\frac{|Z_i-f(u,v; p,m)|^2}{\sigma_i^2}} + 
\frac{(m-m_0)^2}{\sigma_m^2}
\end{equation}
The sum/product can extend over any suitable subset of the visibility 
points. It is useful to be more explicit about the parameters $p$ \& $m$. 
The free parameter $p$ is used to describe the shape of the model, e.g.\ 
the index of the inner part of the broken power-law density distribution. 
The parameter $m$ allows us to include the observational uncertainties,
e.g.\ the $\sim 10$\% uncertainty in the normalization of the models derived
from the Tafalla et al.\ (2003) map, and the $\sim 15$\% uncertainty in the
flux calibration for the BIMA data. This is achieved by allowing the BIMA 
model visibilities to be scaled by a constrained parameter $m$, assumed to 
be a Gaussian random variable with mean $m_0=1.0$ and standard deviation 
$\sigma_m=20$\%. The longer baselines of PdBI do not detect any signal from 
the L694--2 core. These measurements are used to constrain a limit on the 
point source contamination and are not included in the model fitting.

For a given model, we evaluate the best fit parameters by minimizing
the modified $\chi^2$ distribution. Uncertainties in the parameter values 
are analyzed using the Monte Carlo technique known as the {\em bootstrap} 
(Press et al.\ 1992). In brief, the dataset is resampled $n$ times 
(typically $n \sim 200$), each time the fitting process is repeated and the 
best-fit parameters recorded, until the distribution of best-fit parameters 
is well sampled. The width of the distribution provides an estimate of the 
uncertainty in the parameters that best fit the original dataset. 

This numerical approach is made necessary by the non-linear nature of the 
fitting parameters, and is especially useful for this analysis because
small variations in the value of the modified $\chi^2$ as defined above 
may represent surprisingly large variations in fit quality. The nature of
the visibility dataset, comprising a very large number of very low 
signal-to-noise measurements, means that even a model with zero signal will 
on average reproduce each visibility measurement to within its uncertainty.
The resulting shallowness of the $\chi^2$ wells causes two models with 
slightly differing parameters to seem almost equally good despite the fact 
that there is ample signal to distinguish them. The numerical approach
essentially bypasses this complication. The issue could also be circumvented
by binning the visibilities (e.g.\ radially in $(u,v)$ distance), to 
increase the signal-to-noise, and then performing a $\chi^2$ fit to the 
binned values. We avoid this solution because the visibility profiles for 
the various models are falling steeply at the baselines of interest, and 
binning the data inevitably introduces a bias in the fit due to contracting
the range of baseline vectors to a single length at the center of the bin.
We use binning only as a graphical tool in order to demonstrate the fit 
quality of a particular model.

\section{Results and Analysis}

\subsection{Limits on Point Source Flux}
The PdBI dataset covers baselines from 12 to 80~k$\lambda$, with an overall 
rms noise of 0.9~mJy. The visibility data are consistent with the noise. 
The long baselines covered by the PdBI data therefore provide an important 
constraint on the maximum compact component of the flux from the 
L694--2 core, $F_c < 2.7$~mJy (3~$\sigma$). 
For optically thin dust, this limiting flux corresponds 
to an implied (disk) mass limit of 
$M \lesssim 5 \times 10^{-4}$~M$_{\odot} \, (60$~K${}/T_{\mathrm{disk}})$
for an opacity $\kappa_{1.3~\mathrm{mm}} = 0.02$~cm$^2$~g$^{-1}$. This limit
is roughly an order of magnitude lower than the 0.002--0.3~M$_{\odot}$ range
of disk masses observed around T-Tauri stars by Beckwith et al.\ (1990), 
and further demonstrates the ``starless'' nature of the L694--2 core. 

The flux limit allows an estimate of the maximum bolometric luminosity
for any compact component embedded in the L694--2 core. For a simple
estimate, we assume a dust opacity spectral index of unity, and model 
the compact component as a graybody of the form:
\begin{equation}
F_{\nu} \: = \: B_{\nu}(\langle T_{{\rm dust}} \rangle) \,
(1-\exp[-\tau_{\nu}]) \, \Omega_{S} \ ,
\end{equation}
where $B_{\nu}(\langle T_{{\rm dust}} \rangle)$ denotes the Planck function 
at frequency $\nu$ for a mean dust temperature 
$\langle T_{{\rm dust}} \rangle$, $\tau_{\nu}$ is the dust optical depth, 
and $\Omega_{S}$ the solid angle subtended by the source (e.g.\ Beckwith
et al.\ 1990). Since the envelope is entirely optically thin at 1.3~mm, 
the graybody must have a flux that is $\leq 2.7$~mJy at this wavelength. 
For a given mean dust temperature, this constraint fixes the mass of the 
compact component. The only remaining parameter is the radius $R$ of the
component, which essentially identifies the wavelength at which the 
emission becomes optically thick.

For a compact component to remain undetected in our PdBI observations 
implies that its bolometric luminosity is:
\begin{equation}
L_{{\rm bol}} \: \lesssim \: 0.07 {\rm~L}_{\odot} \, (R / 100{\rm~AU})^{0.1}
\: (\langle T_{{\rm disk}} \rangle / 50{\rm~K})^{3.9}
\end{equation}
Note that these power law dependencies are an approximation and are not
accurate for changes in the parameters by factors of more than $\sim 2$.
The dependences on temperature and radius can be understood in the 
following way. The dominant contribution to the luminosity comes from the 
R-J region of the Planck function, in which the emission is optically thin 
(for this typical size scale and temperature, the compact component 
becomes optically thick slightly to the Wien side of the peak in the 
Planck function).  
If the emission from beyond the peak in the Planck function is entirely
negligible, then the luminsosity is proportional to: 
$L \propto \int_{0}^{\nu_{{\rm max}}}{\kappa_{\nu} \nu^2 d\nu}$. 
Note that the flux constraint at 1.3~mm causes there to be no temperature 
dependence in the integrand.
Using the Wien law for $\nu_{{\rm max}} \propto \langle T \rangle$, 
and $\kappa_{\nu} \propto \nu$, then gives 
$L_{{\rm bol}} \propto \langle T \rangle^4$. The contribution of the
optically thick emission causes the actual temperature dependence 
to depart from this relation, and introduces a weak dependence on 
the radius of the compact source.
 
The limit the PdBI observations place on the maximum bolometric luminosity 
of any embedded compact object is substantially lower than the IRAS limit 
for L694--2 of $\sim 0.3$~L$_{\odot}$. Reprocessing of the warm compact 
emission to long wavelengths by the envelope essentially allows a large 
bolometric luminosity to remain hidden below the IRAS detection limit.

\subsection{Density Structure}
The PdBI data constrains the maximum compact component of dust emission,
which allows us to proceed to fit for the envelope structure using the 
BIMA data at shorter baselines. The envelope visibility profiles fall
steeply with increasing baseline. (Note that a flatter emission profile 
leads to a steeper visibility falloff.) At the longer BIMA baselines, 
the visibility amplitudes for the best fit models falls well below the 
signal-to-noise of the entire BIMA dataset. This makes restricting the 
visibility fitting to the shorter baselines advisable, since extending 
the fitting range to longer baseline adds noise but no signal.
Having experimented with various upper limits on the baseline length, we 
found that an upper limit of 10~k$\lambda$ was optimal. Increasing the upper
limit to 15~k$\lambda$, or 20~k$\lambda$ did not change the results but 
produced higher values of the reduced $\chi^2$. The visibility dataset 
to which the models were fit contains 6424 visibility measurements (each 
a 3.45~minute integration) and covers baselines from 5 to 10~k$\lambda$. 
The rms noise in this reduced dataset is 3.3~mJy. Table~2 summarizes the 
results from the $\chi^2$ fits to the four types of model described in 
Section~\ref{sec:models}. Below we describe these results.

{\em Bonnor-Ebert Sphere.}---  the model that best fits the visibility 
data has dimensionless outer radius $\ximax=18 \, {}^{+3}_{-4}$ (Fit~I). 
At the 2~$\sigma$ level this is consistent with the results of the 
extinction study, where a B-E sphere with 
$\ximax=25\pm3$ reproduced the observed color excess. 

{\em Plummer-like model.}--- the embedded model that best fits the 
visibility data has scale radius $R_0=26 \, '' \, {}^{+4}_{-3} \; {}''$,
or $6500^{+1000}_{-750}$~AU (Fit~II). This turn-over radius is 
consistent with the Visser (2000) analysis of SCUBA data, which fit a 
broken power law with a transition to a shallower index within a break 
radius of $\sim 32''$ (8000~AU).

{\em Broken Power law.}--- the model that best  fits the visibility data 
has a power law index in the inner region ($r<30''$) of 
$p=0.9 \, {}^{+0.12}_{-0.16}$ (Fit~III).
The turn-over radius used in this model is similar to that from the
Visser (2000) analysis, and the fitted inner power law index is consistent
with their result of $p=0.8$ (no uncertainties are given), despite their
use of an isothermal dust temperature distribution. Note that the use of 
a smaller turn-over distance results in a lower fitted value of $p$. The 
above result should therefore be construed as a ``mean'' index in the 
$r<30''$ (7500~AU) region, or as an upper limit on the power law index in 
the innermost regions.

{\em Cylinder.}--- the end-on cylindrical model that best fits the 
visibility data has scale height $H=12 \, '' \, {}^{+3.0}_{-1.5} \; {}''$,
or $3000^{+750}_{-400}$~AU (Fit~IV). This is consistent with the results of 
the extinction study, where a slightly tilted cylinder with 
$H=13.5'' \pm 1.5''$ best matched the observed color excess. 

For each of the best fit models, the fitted value of the scaling parameter
is $m \simeq 0.9 \pm 0.05$. That the preferred value is similar for each 
model and not unity suggests that the data prefer a particular slope of
visibility profile that when extrapolated back to zero baseline has 10\% 
lower normalization than the MAMBO flux. This difference is well within
the combined uncertainties of the MAMBO flux measurement and the BIMA 
calibration accuracy (total $\sim 20$\%). 

Figure~1 shows plots of visibility amplitude vs.\ $(u,v)$ distance for
the various best fit models, and models that differ by $\pm 2\, \sigma$ in 
the fitting parameters. The profiles of the best fit models are remarkably 
similar, and we return to this point in Section~\ref{sec:disc}. 
The mean signal in the BIMA dataset used to fit the models is shown on the 
plots (filled circle) with $\pm 2\, \sigma$ error bar. The binning 
introduces an uncertainty (and/or bias) in the appropriate $(u,v)$ distance 
for the data point, that is not present in the model fitting procedure. For 
each type of model, we have adopted the $(u,v)$ distance that provides a 
match with the best-fit model. This is motivated by purely graphical 
reasons, but we note that the mean baseline length weighted by the signal 
to noise of the best fit model (for constant noise on each baseline) 
provides an essentially indistinguishable $(u,v)$ coordinate. However, the 
greater spread of the fitted models compared to the binned data point 
demonstrates the effect of this ``smearing'' of the signal in $(u,v)$ 
distance.

The plots in Figure~1 also show $\pm 2\, \sigma$ confidence intervals
of an azimuthally averaged synthetic visibility profile derived from the
Tafalla et al.\ (2003) MAMBO map of L694--2. The visibility profiles are
computed assuming the MAMBO map to be the true intensity distribution
convolved with a Gaussian beam of $11''$ FWHM. No assessment has been made
for systematic errors in the profile due to chopping and subsequent 
reconstruction of the MAMBO observations. The synthetic confidence intervals 
are extended until the the signal to noise drops below 2 (i.e.\ the 
intervals become consistent with zero signal).

\subsubsection{Systematic Uncertainties}
\label{sec:uncer}
Table~3 lists the main sources of systematic uncertainty along with 
the level of uncertainty they produce in the fitted density parameters
($\ximax$, $r_0$, $p$, \& $H$). The various systematic errors are 
discussed below, in rough order of importance: 

{\em Dust temperature distribution.}--- This is the largest source of 
systematic uncertainty. The temperature distribution that has been assumed 
in the analysis varies from $\sim 8$~K at the center to $\sim 14$~K in the 
outer regions, and it is motivated by physical argument. Nevertheless, 
the profile is probably uncertain at the level of $\pm 2$~K. To study 
the effect of this uncertainty on the inferred density structure, we have 
repeated the analysis for each model, replacing the physical temperature 
distribution with an isothermal distribution at 12~K.
The removal of the central minimum in the temperature profile results in 
a given density model producing a steeper emission profile and 
a shallower visibility profile. The density structure inferred is therefore 
less centrally concentrated than with the physical temperature profile. 
The effect is of a similar magnitude for all of the models: 
the best-fit parameters are changed by close to 2~$\sigma$.

{\em Unresolved compact component.}--- The PdBI data limit the maximum 
contribution to the emission from an unresolved compact component: 
$F_c < 2.7$~mJy (3~$\sigma$). Including a dimmer compact component
in the model fitting essentially reduces the visibility amplitude
that is attributable to the envelope and results in fitted density structures
that are more shallow. For a point source flux of $F_c=2$~mJy, the density
structure parameters are changed by roughly 1~$\sigma$.

{\em Central position.}--- The flatness of the inner density profile and the
lack of a detected compact component make the appropriate central position
for the L694--2 core uncertain. However, the baselines of the BIMA 
visibilities from which the density structure is inferred are short and 
correspond to a fringe size of $\sim 20''$. This means that the results are 
not very sensitive to small changes in the adopted central position (the 
pointing center of the PdBI observations). The signal in the combined 
visibilities peaks at $10$~mJy within $1''$ of the adopted central position 
--- the peak of N$_2$H$^+$(1--0) emission. A shift in the central position 
of $5''$, i.e.\ the shift between the fitted center from the extinction 
study and central position assumed here, leads to a $\sim2$~mJy reduction 
in the binned signal and a 1~$\sigma$ change in the density structure 
parameters, corresponding to shallower density profiles.

{\em Extended structure.}--- A leading source of systematic uncertainty in
the L694--2 extinction study was the effect of the contribution of the
extended structure in which the core appears to be embedded (Harvey et al.\
2003b). Including this additional component to the models lead to inferred 
core density profiles that were significantly more steep, consistent with 
Plummer-like and isothermal cylinder models. In the present study, this 
uncertainty has much smaller effect on inferred structure. Since the 
extended structure must be largely resolved out by the interferometers, 
neglecting the extended structure entirely leads to only a 10\% increase 
in the amplitudes of the visibility profiles of the model cores. Repeating 
the fitting analysis for the two types of model in this context results 
in inferred density profiles that are $\sim 0.5$~$\sigma$ more shallow.

{\em Outer boundary.}--- The outer boundary of the L694--2 core has little
effect in the modeling due to the limited field of view of BIMA and PdBI.
For BIMA, the intensity distribution at the assumed outer edge of the 
core is suppressed by a factor $\sim 0.05$ due to the antenna pattern. 
Moreover, the intensity at the edge is already lower than the intensity at 
the center by a factor of $\sim 500$ (Fit~I). Modifying the outer boundary
therefore has no effect on the shape of the inferred density profile. 
However, for the Bonnor-Ebert sphere (Fit~I), changing the outer radius does
modify the density structure parameter $\ximax$ because this parameter 
describes the dimensionless profile, not the physical profile. For this fit,
the alternative structure parameter, the scale radius 
$r_0=R_{\mathrm{out}}/\ximax$ is unchanged.

{\em Dust opacity spectral index.}--- The dust opacity adds systematic error
via the use of the spectral index to transform the flux normalization from
MAMBO to the frequency at which our observations were made. To some extent,
the appropriate MAMBO frequency itself depends on the spectral index, 
because of the broad bandwidth used in the single dish bolometer 
observations. The spectral index of the dust opacity should lie in the range 
$\beta=$1--2. We have used $\beta=1$ throughout; an alternative extreme, 
$\beta=2$, leads to only a 4\% reduction in the flux normalization, and to 
inferred density structure that is more shallow, but to a degree that is 
negligible in comparison to the other sources of error (both systematic and 
random).

\subsection{Discussion: A Physical Density Distribution for L694--2}
\label{sec:disc}
The fitted physical models demonstrate clearly that the steep
density gradient observed in the $r>30''$ region with extinction
(Harvey et al.\ 2003b) does not continue in the inner region that
could not be probed in that study. This is consistent with the 
modeling of single dish dust emission observations by Visser (2000) 
and Tafalla et al.\ (2003).

The visibility profiles of the best fit models shown in Figure~1 are
remarkably similar to each other, only the power law model differing 
enough to be distinguishable by eye, and then only at long baselines.
This provides an interesting demonstration of the degeneracy of the
various models for starless core density structure. Figure~2 presents
a plot of the best-fit density models. The normalization of the density
assumes an opacity of $\kappa_{{\rm 1.3mm}}=0.02$~cm$^2$~g$^{-1}$, which
is uncertain by a factor of $\sim 5$ or more (Ossenkopf \& Henning 1994). 
At large radii the Bonnor-Ebert and Power-law models differ from the 
Plummer-like and cylinder models due to the fact that the latter models
are assumed to be embedded in an extended uniform distribution of gas.
The Plummer-like and Bonnor-Ebert models are almost identical, and have
central densities that differ by only 5\% 
($n({\rm H}_2) = 1.4 \times 10^5$~cm$^{-3}$ with uncertainty $\sim 50$\%).
The density profile of the end-on cylinder departs significantly from the 
the Bonnor-Ebert and Plummer-like profiles. The profile is more shallow 
with much lower central density 
($n({\rm H}_2) = 5.1 \times 10^4$~cm$^{-3}$ with uncertainty $\sim 30$\%). 
This occurs despite the identical form
of the expressions for the cylinder and Plummer-like models (with 
$R_0=\sqrt{8} H$), and the fact that the cylinder and Bonnor-Ebert 
models both represent a balance between self-gravity and thermal pressure. 
The difference derives from the lower dimensionality of the cylindrical 
model; a given line-of-sight corresponds to a constant ``radius'', and 
hence there is no radial integration that for the spherically symmetric 
models makes the column density profiles more shallow than the density 
profiles. In addition, the extension of the cylinder along the 
line-of-sight means that the densities are correspondingly lower at a 
given radius. The asymmetry of the L694--2 core
viewed in extinction provides a basis for preferring the cylindrical
model over the others. A tilted cylinder with $H=13.5$, 
$L \sin{\phi}=0.14$~pc  ($L \simeq 0.2$--$0.5$, and 
$\phi \simeq 20$--$45^{\circ}$), and central density
($n({\rm H}_2) = 4 \times 10^4$~cm$^{-3}$ embedded in a uniform distribution
of gas with column density 
$N(H+H_2) \sim 6 \times 10^{21} \, (L / 0.5$~pc$)$~cm$^{-2}$, 
successfully reproduces the dust emission visibility profile, as 
well as the profile and asymmetry of the dust extinction map. The 
instability of the cylindrical model along its axis is also consistent
with the inward motions in L694--2 inferred from molecular 
spectral lines (Lee, Myers \& Tafalla 2001).

As already noted, the inner power law index of the fitted broken 
power law model ($p=0.9 \, {}^{+0.12}_{-0.16}$, Fit~II) may be construed 
both as a ``mean'' index in the $r<30''$ region, and as an upper limit on 
the index in the innermost regions (due to the trade-off between turn-over 
radius and power law index in the inferred visibility profile). The
former is illustrated by the fact that the radially averaged ``mean'' 
effective power law index for the best-fit Plummer-like model is 
$\bar{p}=0.86$ over the range in radius $0 \leq r \leq R_0$ (with 
$R_0=26''$). The profile is consistent with the Visser (2000) study
who found that a similar index ($p=0.8$) and turn-over radius ($R_0=32''$)
reproduced the SCUBA map.
The index in the inner region is much less than that observed in the 
outer envelope and shows that the density distribution in the inner region
of the L694--2 core has clearly not relaxed fully, along the lines
assumed in the inside-out collapse of Shu (1977), despite the presence
of extended inward motion in the gas (Lee et al.\ 2001).
These inward motions may reflect the contraction of a prolate core 
along its major axis (Harvey et al.\ 2003b).

\section{Summary}

We present a study of the density distribution of the candidate 
contracting starless core L694--2 using high resolution 1.3~mm 
dust continuum observations from BIMA and the IRAM PdBI. In summary:

\begin{enumerate}
\item The PdBI visibility data span baselines from 12 to 80~k$\lambda$
(spatial scales from $\sim 500$ to $\sim 5000$~AU) and do not detect 
any significant emission from the core. This provides a stringent constraint 
on the maximum point source flux: $F_c < 2.7$~mJy (3~$\sigma$). 
This flux limit corresponds to a very small disk mass, 
$M \lesssim 5 \times 10^{-4}$~M$_{\odot} \, (60$~K${}/T_{\mathrm{disk}})$,
and bolsters the ``starless'' interpretation of the L694--2 core.

\item The BIMA visibility data in the baseline range 5 to 10~k$\lambda$
(spatial scales $\sim 10^3$--$10^4$~AU) detect emission from the core 
and are used to constrain models for the density structure. That no signal 
is evident on baselines beyond 10~k$\lambda$ confirms the flat emission
profile in the inner regions, and a significant change in behavior from 
the steep density profile for $r\gtrsim 8000$~AU inferred from near infrared 
extinction of background stars (Harvey et al.\ 2003b).
We fit four types of model for starless core density structure that 
can match the steep density gradient in the outer regions inferred from 
extinction. We adopt a temperature distribution that decreases in the inner
regions due to shielding from the ISRF (Evans et al.\ 2001).
The best fit Bonnor-Ebert sphere has dimensionless outer radius
$\ximax=18 \, {}^{+3}_{-4}$ (1~$\sigma$). The best fit Plummer-like model
has turn-over radius $R_0=26 \, '' \, {}^{+4}_{-3} \; {}''$ 
($6500^{+1000}_{-750}$~AU). The best fit
broken power law model has index $p=0.9 \, {}^{+0.12}_{-0.16}$ (this model
uses the maximum allowable turn-over radius of $R_0=30''$ (7500~AU) and 
provides an upper limit on the index at the center of the core). 
The best fit end-on isothermal cylinder has scale height 
$H=12 \, '' \, {}^{+3.0}_{-1.5} \; {}''$ ($3000^{+750}_{-400}$~AU).

\item We consider the effects of various sources of systematic uncertainty
on the derived density structure. The largest uncertainty comes from the
dust temperature distribution; assuming an isothermal core modifies the
best fit parameters by $\sim 2\, \sigma$, and makes the inferred density
profiles more shallow. A possible weak compact component to the emission
($F\sim 2$~mJy) and the appropriate central position 
($\delta \theta \lesssim 4''$) are the main remaining sources of 
uncertainty. Both of these uncertainties lead to $\sim 1 \, \sigma$ changes 
in the inferred model parameters, again to shallower density profiles.

\item The density profiles of the various best-fit models are nearly 
indistinguishable at baselines shorter than $\sim 10$~k$\lambda$. 
There is a strong degeneracy between the models for starless core density
structure. The two physical models (B-E sphere and isothermal cylinder) 
represent a balance between self-gravity and thermal pressure, and therefore 
they produce similar profiles. The asymmetry of the L694--2 core evident
in the extinction map may provide a basis for favoring the cylindrical 
model. A cylinder slightly tilted to the line-of-sight 
with $H=13.5''$, $L \sin{\phi}=0.14$~pc ($L\simeq 0.2$--$0.5$, and 
$\phi \simeq 20$--$45^{\circ}$) and central density
($n({\rm H}_2) = 4 \times 10^4$~cm$^{-3}$ embedded in a uniform distribution
of gas with column density 
$N(H+H_2) \sim 6 \times 10^{21} \, (L/0.5$~pc$)$~cm$^{-2}$ reproduces
both the dust emission and extinction measurements.
\end{enumerate}

\acknowledgements
We acknowledge the IRAM and BIMA staff for carrying out the observations.
We are especially grateful to Roberto Neri and Jerome Pety for coordinating 
remote reduction of the IRAM PdBI observations.

\clearpage

\begin{deluxetable}{lll}
\tablenum{1}
\tabletypesize{\footnotesize}
\tablewidth{0pt}
\tablecolumns{3}
\tablecaption{Summary of instrumental parameters}
\tablehead{\colhead{Parameter} & \colhead{PdBI Value} & 
\colhead{BIMA Value}}
\startdata
Observation dates & 
 2001 Nov.\ 20; &
 2001 Sep.\ 22; Sep.\ 29;\\
 & 2002 Apr.\ 03 & 2002 Mar.\ 31 Jun.\ 08; Sep.\ 09\\
Configuration & 
 D (four ants.); D (six ants.) &
 D; D; C; D; D (nine ants.)\\ 
Pointing center (J2000) & 
 $19^{h}41^{m}04\farcs44$, $+10^{\circ}57'00\farcs9$ &
 $19^{h}41^{m}04\farcs32$, $+10^{\circ}57'06\farcs1$\\
Observing frequency & 
 231.3~GHz &
 231.3~GHz\\
Phase calibrators & 
 J1751+096, J1925+211 &
 J1925+211\\
Bandpass calibrator & 
 3C~345; 3C~273 &
 Uranus\\
Flux calibrator & 
 MWC~349 & 
 Uranus\\
Primary beam FWHM & 
 $22''$ &
 $50''$\\
Bandwidth & 
 480~MHz &
 700~MHz\\
RMS & 0.9~mJy & 1.6~mJy\\
\enddata
\end{deluxetable}

\begin{deluxetable}{lllcc}
\tablenum{2}
\tabletypesize{\footnotesize}
\tablewidth{0pt}
\tablecolumns{5}
\tablecaption{Summary of the Density Model Fits \label{tab:fits}}
\tablehead{ \colhead{Fit} & \colhead{Density Model} & 
\colhead{Fitted Model Parameter} & \colhead{Calibration scaling} & 
\colhead{$\chi^2_\nu$}}
\startdata
I & Bonnor-Ebert & $\ximax=18 \, {}^{+3}_{-4}$ & 
$M=0.90 \pm 0.04$ & 1.2520 \\
II & Plummer-like & $R_0=26 \, '' \, {}^{+4}_{-3} \; {}''$ & 
$M=0.90 \pm 0.05$ & 1.2521\\
III & Broken Power Law & $p=0.9 \, {}^{+0.12}_{-0.16}$ & 
$M=0.92 \pm 0.04$ & 1.2519 \\
IV & Cylinder & $H=12 \, '' \, {}^{+3.0}_{-1.5} \; {}''$ & 
$M=0.90 \pm 0.05$ & 1.2520\\
\enddata
\end{deluxetable}

\begin{deluxetable}{llc}
\tablenum{3}
\tabletypesize{\footnotesize}
\tablewidth{0pt}
\tablecolumns{3}
\tablecaption{Summary of systematic uncertainties in fitted parameters
\label{tab:errors}}
\tablehead{\colhead{Model Assumption} & \colhead{Variation} 
& \colhead{Resulting Systematic Error ($\sigma$) \tablenotemark{1}}}
\startdata
Temperature distribution & 
$T_d=T_{\mathrm{phys}}(r) \longrightarrow 12$~K & $\sim 2$\\
Central point source flux & $F \sim 2$~mJy & $\sim 1$\\
Central position & $\delta \theta \lesssim 4''$ & $\sim 1$\\
Neglect extended structure~\tablenotemark{2} & $F_0(r<30'')=80$~mJy 
$\longrightarrow 0$ & $\sim 0.5$\\
Outer boundary of L694--2~\tablenotemark{3} & 
$\delta R_{\mathrm{out}} / R_{\mathrm{out}} \lesssim 50$\% & $\ll 1$\\
Dust opacity spectral index & $\beta=1 \longrightarrow 2$ & $\ll 1$\\
\enddata
\tablenotetext{1}{Systematic error in the fitted model parameter given in 
units of $\sigma$ the random error in the parameter listed in Table~2. 
Note that each quoted systematic error corresponds to a {\em shallowing} 
of the inferred density profile.}
\tablenotetext{2}{Applies to Plummer-like and Cylinder models only}
\tablenotetext{3}{Does not apply to B-E sphere; for the B-E model, 
increasing the outer radius increases $\ximax$ in proportion to 
$R_{\mathrm{out}}$, keeping constant scale radius $r_0$}
\end{deluxetable}

\clearpage
\begin{figure}
\figurenum{1}
\epsscale{0.5}
\plotone{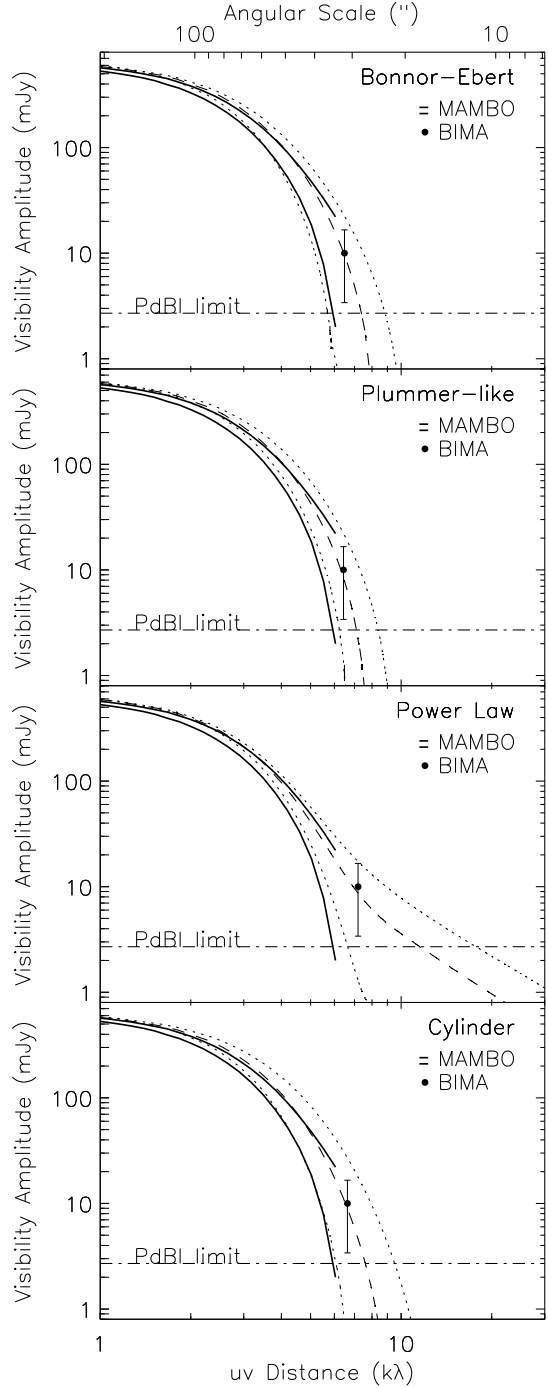}
\caption{Visibility amplitude vs.\ $(u, v)$ distance for L694--2 at 1.3~mm,
for the best-fitting models (dashed lines) and models that deviate by 
$\pm 2 \, \sigma$ in the fitting parameters (dotted lines). The binned
BIMA visibility amplitude in the baseline range 5 to 10~k$\lambda$ is also
shown (circle), with a $\pm 2 \, \sigma$ error bar. The PdBI limit on a 
point-like component ($>12$~k$\lambda$) is shown as a horizontal (dash-dot) 
line.  The solid lines represent $\pm 2 \, \sigma$ intervals of the 
azimuthally averaged synthetic visibility profile derived from the MAMBO 
map. See text for further explanation.}
\end{figure}

\clearpage
\begin{figure}
\figurenum{2}
\epsscale{0.8}
\plotone{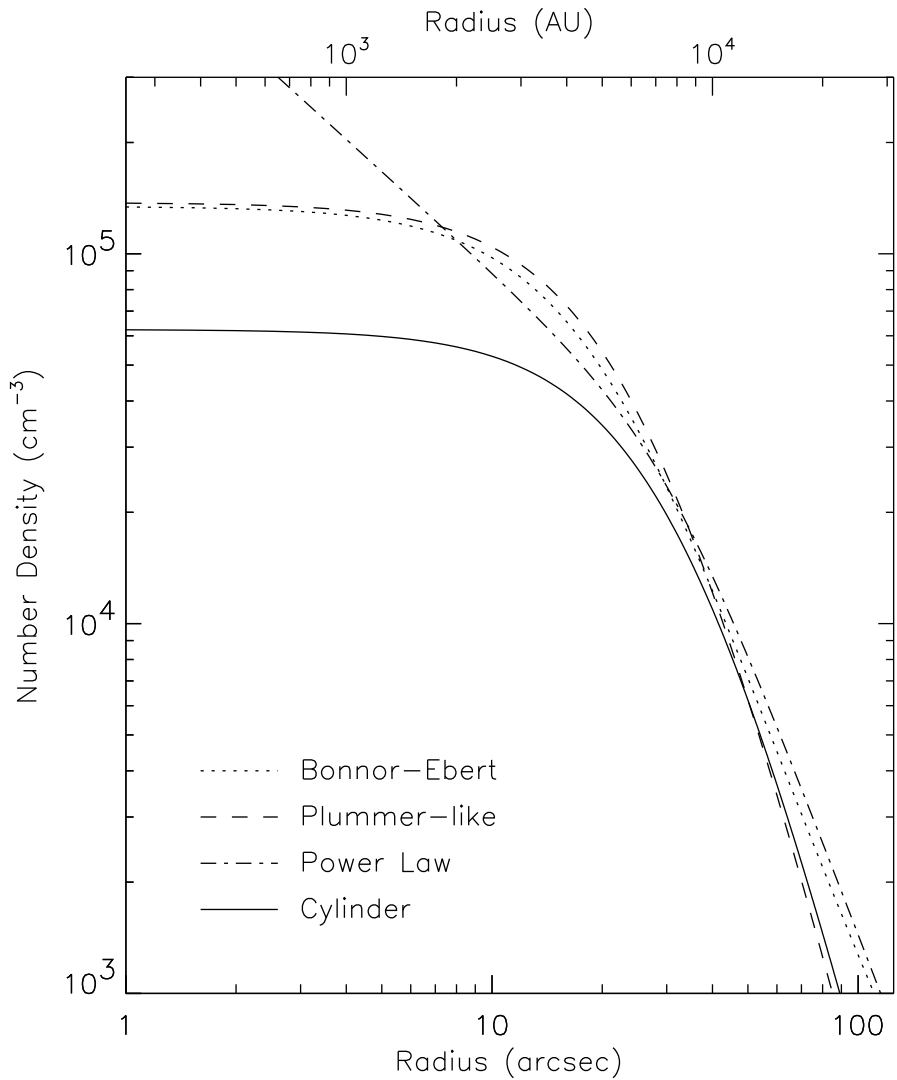}
\caption{Number density of molecular hydrogen vs.\ radius for the various
best fit models of L694--2. The number density calculation assumes a dust
opacity of $\kappa_{{\rm 1.3mm}}=0.02$~cm$^2$~g$^{-1}$, a mean
molecular weight of $\mu=2.29$, and a Hydrogen mass fraction 
$X_{{\rm H}}=0.73$. The best-fit Bonnor-Ebert ({\em dotted}) and Plummer-like
({\em dashed}) profiles are nearly identical. The cylindrical model 
({\em solid}) has density a factor $\sim 2$ lower due to the extension
of the cylinder along the line-of-sight.}
\end{figure}

\end{document}